\documentclass[twocolumn,prl,showpacs,aps]{revtex4}

\usepackage{amssymb}
\usepackage{amsmath}
\usepackage{latexsym}
\usepackage{hyperref}

\begin{document}
\title{Large Extra Dimensions and Holography}
\author{Chao Cao}\email{ccldyq@gmail.com}
\author{Yi-Xin Chen}\email{yxchen@zimp.zju.edu.cn}
\affiliation{Zhejiang Institute of Modern Physics, Zhejiang
University, Hangzhou 310027, China}

\begin{abstract}
The holographic principle asserts that the entropy of a system
cannot exceed its boundary area in Planck units. However,
conventional quantum field theory fails to describe such systems. In
this Letter, we assume the existence of large $n$ extra dimensions
and propose a relationship between UV and IR cutoffs in this case.
We find that if $n=2$, this effective field theory could be a good
description of holographic systems. If these extra dimensions are
detected in future experiments, it will help to prove the validity
of the holographic principle. We also discuss implications for the
cosmological constant problem.

\end{abstract}
 \maketitle

\emph{Introduction.---}One of the most important mysteries in modern
physics is the explanation of black hole entropy. With a
semiclassical analysis Bekenstein \cite{Bekenstein} and Hawking
\cite{Hawking} argued that a black hole with radius $L$ and horizon
area $A$ radiates and has entropy
\begin{equation} \label{eq:1}
S_{\rm BH}=\frac{A}{4l_p^2}\sim (\frac{L}{l_p})^2,
\end{equation}
where $l_p=\sqrt{\hbar G/c^3}$ is the Planck length. This
non-extensive nature has led to the ``holographic principle"
\cite{Hooft, Susskind}, which asserts that the number of degrees of
freedom (DOF) inside a bounded region should not exceed the surface
area $A$.

Although the holographic principle remains a conjecture, it has
received some strong evidences during recent years. One famous
example is the Ads/CFT correspondence (anti-de-Sitter
space/conformal field theory correspondence) \cite{Ads}. However,
this entropy bound has posed a challenge to the local quantum field
theory (LQFT) as the underlying theory of nature \cite{Hooft}. For a
LQFT in a box of size $L$ with UV cutoff $\Lambda$, the entropy $S$
scales extensively, $S\sim (\frac{L\Lambda}{\hbar})^3$, which will
be obviously larger than $S_{\rm BH}$. The failure of LQFT in
describing a holographic system seems to be a disaster, since the
effective field theory is so successful in describing elementary
particles that it is still one of the most useful tools in modern
physics until now. To avoid this difficulty, Cohen et al.
\cite{Cohen} proposed a constraint on the UV cutoff. By excluding
all states of the system that already have collapsed to a black
hole, they got a much stronger limit
\begin{equation} \label{eq:2}
L^3\Lambda^4\lesssim LM_p^2\hbar^2c^2,
\end{equation}
where $M_p=\sqrt{\hbar c/G}$ is the Planck mass. Then the UV cutoff
scales like $L^{-1/2}$. This entropy bound is less than the
holographic entropy
\begin{equation} \label{eq:3}
S\lesssim (\frac{L}{l_p})^{\frac{3}{2}}\sim A^{\frac{3}{4}}.
\end{equation}
The $A^{3/4}$ bound was firstly obtained by G. 't Hooft in
\cite{Hooft}. It can also be derived by a strict counting of all the
non-gravitational collapse quantum states \cite{chen}. Nevertheless,
this bound is far more restrictive. There is obviously a gap from
this entropy bound to the holographic entropy, which would cause a
huge numerical differences from $10^{90}$ to $10^{120}$ for our
present universe. What is the DOF filling up the gap is still a
question to be addressed.

We should note that the work \cite{Cohen} was done in the
4-dimensional spacetime, however, many recent works suggest the
existence of extra dimensions \cite{Nima, Nima2, RS, string}. From
the 4-dimensional point of view, the states from extra dimensions
should give additional DOF. This may help to fill the gap between
the two kinds of entropy (\ref{eq:1}) and (\ref{eq:3}). In this
Letter, we try to extend the UV-IR relation by considering $n$ extra
dimensions at a millimeter, which comes from the hierarchy problem
\cite{Nima}. The entropy bound of this new field theory obeys a
strict area-law when $n=2$. Moreover, these two extra dimensions
might be detected by the gravitational inverse-square law test and
the LHC experiments in the future. If this is true, it will give
strong support to the holographic principle.

One other success in Cohen et al.'s work is that the UV-IR related
model (\ref{eq:2}) also has an advantage of solving the celebrated
cosmological constant (CC) problem. In this Letter, we find that our
new model can also estimate the vacuum energy and eliminate the need
for fine-tuning.

\emph{UV-IR relation in $4+n$ dimensions.---}In 1998, an elegant
framework was proposed by Arkani-Hamed et al. (ADD) \cite{Nima} for
solving the hierarchy problem: the Planck scale
$M_{p}\sim10^{19}{\rm GeV}/c^2$, which reflects the strength of
gravity, is 16 orders of magnitude larger than the electroweak scale
$M_{EW}\sim1{\rm TeV}/c^2$. Introducing $n$ extra compact spatial
dimensions with radius $R$, the traditional Planck scale $M_p$ is
only an effective energy scale derived from the fundamental
$(4+n)$-dimensional one, $M_*$. By applying Gauss' law, one finds
\begin{equation} \label{eq:4}
M_p^2\sim (\frac{Rc}{\hbar})^n M_*^{n+2}.
\end{equation}
Setting $M_*\sim M_{EW}\sim1{\rm TeV}/c^2$ yields
\begin{equation} \label{eq:5}
R\sim10^{\frac{30}{n}-19}{\rm m}.
\end{equation}
The $n=1$ case is subsequently excluded, since $R\sim 10^{11}{\rm
m}$ has to be of the order of radius of the solar system. From
\cite{Nima2}, $M_*\sim 1{\rm TeV}/c^2$ for $3\leq n\leq 6$, while
for the $n=2$ case, $M_*$ takes an astrophysically preferred value
$\sim 1{\rm TeV}/c^2-10{\rm TeV}/c^2$, and then $R\sim 10^{-5}{\rm
m}-10^{-3}{\rm m}$. On the other hand, the success of the Standard
Electroweak Model (SM) up to $\sim 100{\rm GeV}$ implies the SM
fields  are localized to the 4-dimensional submanifold of thickness
$M_{EW}^{-1}$ in the extra $n$ dimensions. The only fields freely
propagating in the $(4+n)$-dimensional spacetime are gravitons. Such
a framework can be accommodated in string theory \cite{string}. The
visible universe is in fact a D3-brane extending over three spatial
dimensions, the gauge particles, as the endpoints of open strings,
are attached to the D-brane, while the gravitons, made of closed
strings, can freely exist in the bulk. We will not sense additional
dimensions except via their modification of the gravitational force
law.

Now we explain how the existence of extra dimensions affects the UV
cutoff. First, we determine the proper mass expression for the
$(4+n)$-dimensional black hole with radius $L$ larger than the size
$R$ of the extra dimensions by virtue of a gedanken experiments
following Laplace \cite{Laplace}. The escape velocity of the black
hole should be the speed of light, i.e., the kinetic energy of a
particle moving at the speed of light should be equal to the
magnitude of its gravitational potential energy while this particle
is placed at the black hole horizon. For the $(4+n)$-dimensional
black hole with a horizon radius $L$ on the 3-brane and mass $M$, we
have
\begin{equation} \label{eq:14}
V(L)\sim \frac{G_nMm}{R^nL}\sim\frac{mc^2}{2},
\end{equation}
where $G_n=\hbar^{n+1}c^{1-n}/M_*^{n+2}$ is $(4+n)$-dimensional
Newton's constant. From Eq. (\ref{eq:14}) we get the expression for
black hole mass is $M \sim R^n L M_*^{n+2}\hbar^{-n-1}c^{n+1}$. Now
consider a homogeneous LQFT system, the non-gravitational collapse
condition is written as
\begin{equation} \label{eq:6}
R^n L^3 \Lambda^{n+4}\lesssim R^n L M_*^{n+2}\hbar^2c^{n+2},
\end{equation}
which gives a new UV-IR relation. In order to prevent this system
collapsing, we should exclude all states whose momentum is higher
than $\Lambda_{\rm max}\sim L^{-2/(n+4)}$ on the 3-brane. However,
we don't need to take this limit in the extra $n$ dimensions, since
(i) the Poincare invariance is now localized on the 3-brane, the
effective UV cutoff taking in 4 dimensions will not affect the
$n$-dimensional compact space, (ii) due to Gauss' law, the change of
energy density along the transverse $n$ extra directions will not
affect the longitudinal potential, unless the 4-dimensional density
goes beyond the critical value, we will never see the gravitational
collapse on the brane. So the UV cutoff in extra dimensions can be
naturally up to the fundamental scale $M_*c$. Moreover, the compact
extra dimensions cause quantized momentum $p=N\hbar/R, N\in Z$, in
the 4D point of view this is equivalent to a tower of Kaluza-Klein
(KK) states. At energies small compared to the KK scale
$\Lambda_{KK}\sim \hbar/R$, these KK states can not be excited, and
this low energy physics is effectively four-dimensional. Thus the
UV-IR relationship remains Eq. (\ref{eq:2}). Only at energies above
$\Lambda_{KK}$, the relation (\ref{eq:6}) is applicable. To see this
more explicitly, define $\Lambda_4\equiv(\frac{M_p\hbar
c}{L})^{1/2}$ and
$\Lambda_*\equiv(\frac{M_*^{n+2}\hbar^2c^{n+2}}{L^2})^{1/(4+n)}$,
which are the bounds of UV cutoff respectively come from Eq.
(\ref{eq:2}) and (\ref{eq:6}). Combine with the relationship
(\ref{eq:4}), we can get
\begin{equation} \label{eq:7}
\Lambda_4\gtrsim \Lambda_{KK}\Leftrightarrow
\Lambda_4\gtrsim\Lambda_* \gtrsim\Lambda_{KK}.
\end{equation}
So once the LQFT cutoff reaches $\Lambda_{KK}$, the system will
collapse into a $(4+n)$-dimensional black hole before it converts
into a 4-dimensional one. The UV cutoff is then lowered down as in
Eq. (\ref{eq:6}).

While $L\ll R$, we can directly use the result of higher dimensional
Schwarzschild black holes (some properties of these small black
holes were discussed in \cite{BH}, they are bigger, colder, and
longer-lived than a usual 4D black hole of the same mass). This
``sub-millimeter" UV-IR relationship is
\begin{equation} \label{eq:8}
L^{n+3} \Lambda^{n+4}\lesssim L^{n+1} M_*^{n+2}\hbar^2c^{n+2},
\end{equation}
same form as (\ref{eq:6}) (maybe with some different coefficients,
this is because black holes with $L>R$ are not spherical). We also
see that $\Lambda_*$ here is obviously higher than $\Lambda_{KK}$,
Eq. (\ref{eq:2}) is not suitable in this case.

We emphasize that the UV and IR cutoffs here are global definitions
in the whole box, i.e., depend on the average energy density bounds.
In a local region of this box, there could also exist higher
density, with lower average density in the rest region to make the
total energy not enough to form a black hole. However, for a system
with definite energy, the uniform state with same energy density
everywhere has the largest entropy \cite{footnote1}, so we can just
consider this situation for simple and give a realistic entropy
bound. In fact, by using uncertainty relationship $L\Lambda\gtrsim
\hbar$ and Eqs. (\ref{eq:6}), (\ref{eq:8}), natural local cutoffs on
$L$ and $\Lambda$ are $L_H\gtrsim L\gtrsim l_*$ and $\Lambda\lesssim
M_*c$, where $L_H$ is current horizon scale and $l_*$ is the
$(4+n)$-dimensional Planck length. These bounds, above which quantum
gravity effects would modify the semiclassical picture, are the real
cutoffs we use in the usual LQFT calculation \cite{footnote2}.
Moreover, since the UV cutoff in extra dimensions can be up to
$M_*c$, we can define KK species number
\begin{equation}\label{eq:9}
N_*\equiv(\frac{M_*c}{\Lambda_{KK}})^n=(\frac{RcM_*}{\hbar})^n,
\end{equation}
then by using Eq. (\ref{eq:4}), we have $N_*\Lambda^2\lesssim
(M_pc)^2$, which is coincident with to the large $N$ bound proposed
by G.Veneziano \cite{N bound1}, Dvali and Redi \cite{N bound}. Both
papers introduce a large number $N$ of particle species. \cite{N
bound1} considered the contribution of gravity and gauge loop, while
\cite{N bound} considered the evaporation process of a macroscopic
black hole carrying the maximal $N$-units of the discrete charges.
Here we derive the $N$ bound in a different way, from  the existence
of large extra dimensions and uncertainty relationship.

\emph{CC problem.---} In conventional LQFT, the quantum contribution
to the vacuum energy density computed in perturbation theory is
$\sim(100 {\rm GeV})^4$, while the discovery of dark energy suggests
a very small vacuum energy density $\sim(10^{-3}{\rm eV})^4$. This
discrepancy is traditionally explained by unknown physics either at
high energies or the anthropic principle noted by S. Weinberg
\cite{Weinberg}.

However, according to Cohen et al.'s theory, the reason for
conventional fine-tuning CC problem is that we did the perturbative
computation of the quantum correction of the the vacuum energy
density without the consideration of an IR limitation to the LQFT.
While taking $\Lambda\sim L^{-1/2}$ into account and choosing $L$ as
the current universe size, the resulting quantum energy density will
be effectively lowed down to $\Lambda_{\rm CC}^4\sim(10^{-3}{\rm
eV}/c)^4$, in full agreement with the current measurement of
cosmological constant.

Now there exist $n$ extra dimensions, as we discussed in Sec. II,
the choice of UV-IR relation mainly depends on the KK scale. For
$n\geq3$, according to Eq. (\ref{eq:5}), $R<10^{-8}{\rm m}$, the
corresponding KK scale is $\Lambda_{KK}\gtrsim 10{\rm eV}/c\gg
\Lambda_{\rm CC}$, this low energy is not enough to excite KK
states, so we should use Eq. (\ref{eq:2}) to compute the effective
vacuum energy density and get the same result as Cohen et al.'s.

However, for $n=2$, $R\sim 10^{-5}{\rm m}-10^{-3}{\rm m}$ and
$\Lambda_{KK}\sim 10^{-4}{\rm eV}/c-10^{-2}{\rm eV}/c$, with respect
to the chosen value of $M_*\sim 1{\rm TeV}/c^2-10{\rm TeV}/c^2$. Now
$\Lambda_{KK}$ is around $\Lambda_4$, by using Eq. (\ref{eq:6}), the
resulting UV cutoff is also $\Lambda_* \sim 10^{-3}{\rm eV}/c$. We
see that the existence of two ``large" extra dimensions also fits
current observation. More encouragingly, since most SM particles are
trapped on the 3-brane, this modified result is mainly vacuum energy
contribution. In fact, Y. Aghababaie et al. \cite{Burgess} have
proposed a explicit six-dimensional supergravity model, in which the
effective 4D cosmological constant can self-tune to current
observation if the radius of extra dimensions is sub-millimeter in
size.

\emph{Entropy bound.---}Consider a 4D system with UV-cutoff
$\Lambda$ in a box of size $L$, let us discretize this space to
lattices, the minimal cell scale should be $\sim \hbar/\Lambda$.
Assume that there is one oscillator per cell and each has a finite
number of states, $m$, then the total number of independent quantum
states in this specified region is $\sim m^{(L\Lambda/\hbar)^3}$.
The entropy defined as the logarithm of the number of microstates is
now $S\sim L^3\Lambda^3/\hbar^3$, seems to scale extensively, and
much exceeds the holographic bound (\ref{eq:1}). However when Eq.
(\ref{eq:2}) is near saturation, the entropy becomes $S\lesssim
(L/l_p)^{3/2}$, more restrictive than Eq. (\ref{eq:1}). A
conventional explanation of this bound is that we are only
explicitly considering those states that can be described by usual
quantum field theory. This is confusing. We impose a relationship
between UV and IR cutoffs in order to make the LQFT still be good in
describing nature, but meanwhile we also require another unknown
theory to fill the missing DOF, against the famous Occam's razor.

However, in the $(4+n)$-dimensional frameworks, from 4D point of
view, for each particle on the 3-brane world there is a tower of KK
excitations, so the total entropy is the 4D one multiplied by the
number of KK species $N_*$
\begin{equation}\label{eq:10}
S\sim N_*(\frac{L\Lambda}{\hbar})^3.
\end{equation}
Substituting Eqs. (\ref{eq:6}), (\ref{eq:8}) and (\ref{eq:9}) into
(\ref{eq:10}), we can get
\begin{equation}\label{eq:11}
S\lesssim
(\frac{L}{l_p^{\frac{2}{n+2}}R^{\frac{n}{n+2}}})^{\frac{n-2}{n+4}}(\frac{L}{l_p})^2.
\end{equation}
When $n=0$, there is no extra dimensions, the entropy bound
$(\frac{L}{l_p})^{3/2}$ and the corresponding UV-IR relation are
still the same as these obtained by Cohen et al. \cite{Cohen}. While
for $n=2$, this entropy bound has become $(L/l_p)^2$, satisfying the
4-dimensional area law. Moreover, as we have discussed above, in the
$n=2$ case, the IR-UV relation (\ref{eq:6}) is applicable up to
current universe size $L_H$, while (\ref{eq:8}) is applicable down
to the fundamental Planck length $l_*$, so this bound can always be
achieved when the system begins to collapse into a black hole.

Now the question is how to explain the entropy bound, since we
derive Eq. (\ref{eq:11}) from a 6D theory, why it represents the
4-dimensional holographic bound? In fact, by using the relation
$l_p\sim\hbar/(M_pc)$, $l_*\sim\hbar/(M_*c)$ and Eq. (\ref{eq:4}),
we have $l_p^2\sim l_*^{n+2}/R^{n}$. Then for $n=2$, one can get
\begin{equation} \label{eq:12}
(\frac{L}{l_p})^2\sim\frac{L^2R^2}{l_*^{4}}.
\end{equation}
We see that this entropy bound is also proportional to the area of
the 6D region, but in different Planck units. This just satisfies
the 6-dimensional holographic bound. The information of the whole 6D
world is stored on its surface, a 4D hologram, while this 4D
hologram can be projected to a much denser and smaller 2D hologram,
which is just the surface of our 4-dimensional world.

It is straightforward to generalize our above proposal to $d+n$
dimensions, with a $d$-dimensional braneworld. In this case, the
UV-IR relation becomes
\begin{equation} \label{eq:13}
R^n L^{d-1} \Lambda^{n+d}\lesssim R^n L^{d-3}
M_*^{n+d-2}\hbar^2c^{n+d-2}.
\end{equation}
In order to get a holographic entropy bound, one should require that
$n=d-2$.

Although the new IR-UV relationship based on two extra dimensions
can give a good description to holographic bound, one should note
that $R$ is very marginal for $n=2$. When we take $M_*\sim 1{\rm
TeV}/c^2$, $R\sim 1 {\rm mm}$ is just the region that present
experimental measurements of gravitational $1/r^2$ law can probe. In
fact, one such test in 2000 \cite{gausslaw} has suggested a
unification scale of $M_*\gtrsim 3.5{\rm TeV}/c^2$. We hope the
gravitational force law could change from $1/r^2$ to $1/r^4$ on
distances $\sim 10^{-5}{\rm m }\sim 10^{-4}{\rm m}$ by new
experiments in the future.

The effects of large extra dimensions might also be detected at the
LHC,  a pp collider with a cm energy of 14 TeV. First, one generic
feature of extra dimensional models is the occurrence of KK modes.
These KK excitations associated with the SM gauge bosons like $W$,
$Z$ and so on might be directly produced at the LHC
\cite{KKdectect}. If such states  really exist, one can determine
not only  the unification scale $M_*$, but also the number of extra
dimensions. Second, if the fundamental planck scale goes down to
$\sim {\rm TeV}$, small black holes will be produced in large
numbers, and the testing of the Hawking radiation will help to
determine the number of compact extra dimensions and the scale of
quantum gravity \cite{BHdectect}. Moreover, extra dimensions also
offer new possibilities for supersymmetry (SUSY) breaking. The
lightest SUSY particles should be in the expected discovery range of
the LHC.

\emph{Conclusions.---}There exist three ``hierarchy" problem in
modern physics. (i) Why the weak force is $10^{32}$ times stronger
than gravity, this problem can be solved by SUSY, technicolor or
various extra dimensions models such as ADD model \cite{Nima} and
Randall-Sundrum model \cite{RS}. (ii) Why current accelerating
universe indicates a tiny cosmological constant which is much
smaller than the vacuum energy predicted by most quantum field
theories. This famous CC problem can be explained by the anthropic
principle \cite{Weinberg} or a UV-IR related LQFT \cite{Cohen}.
(iii) The holographic bound which obeys the area-law is either much
lower than the conventional LQFT entropy bound, or much higher than
the 4-dimensional UV-IR related LQFT entropy bound.

In order to solve the third problem, we introduce $n$ extra compact
spatial dimensions, which was first proposed in ADD model, into the
UV-IR related LQFT.  For $n=2$, we can exactly retrieve a
holographic entropy bound. Just as in \eqref{eq:4} the effect of
extra dimensions lowers the 4-dimensional Planck scale $M_p$ down to
the electroweak scale $M_{EW}$, in \eqref{eq:10} and \eqref{eq:11}
it also magnifies the LQFT entropy bound $A^{3/4}$ to a holographic
one, while it is easy to identify the contribution of KK states to
the DOF counting. Actually our approach has provided an opportunity
to settle the three hierarchy problems in a unified framework. We
point out that $n=2$ is a dangerous but exciting case. We hope the
gravitational force law test and LHC can help to search these large
dimensions. Once the existence of these two dimensions is really
true, one may infer that the holographic principle is generally
valid for all the systems including black holes and these LQFT
systems.

We thank J. L. Li and Y. Xiao for useful discussions. This work is
supported in part by the NNSF of China Grant No. 90503009, No.
10775116, and 973 Program Grant No. 2005CB724508.


\begin{thebibliography}{9}
\bibitem{footnote1} This state actually obeys an equiprobability distribution, which of course has
the largest entropy. For example, consider a box, separate it into
two regions A and B, acquire $\rho_A V_A+\rho_B V_B\leq M$. Then the
total entropy in the box $S\sim \rho_A^{\alpha} V_A+\rho_B^{\alpha}
V_B$ ($\alpha$ depends on the number of dimensions) takes the
maximal value just when $\rho_A=\rho_B=M/(V_A+V_B)$.
\bibitem{footnote2} Note that in local processes, the momentum of course varies
from $\hbar /L$ to $M_*c$, although the average energy is confined
below $\Lambda_*c$.
\end{thebibliography}
\end{document}